\documentclass[twocolumn,pre,showpacs]{revtex4}
\usepackage{graphicx}
\usepackage{amsmath}
\usepackage{amsfonts}
\usepackage{amssymb}
\usepackage{color}
\usepackage{float}
\usepackage{mathrsfs}

\begin{document}
\title{Growing of integrable turbulence}

\author{D.\,S.~Agafontsev$^{1,2}$}
\email{dmitrij@itp.ac.ru}
\author{V.\,E.~Zakharov$^{2,3}$}

\affiliation{\textit{$^1$ Shirshov Institute of Oceanology of RAS, 117997 Moscow, Russia.\\
$^2$ Skolkovo Institute of Science and Technology, 121205 Moscow, Russia.\\
$^3$ Department of Mathematics, University of Arizona, 857201 Tucson, AZ, USA.}}

\begin{abstract}
We study numerically the integrable turbulence in the framework of the focusing one-dimensional nonlinear Schr{\"o}dinger equation using a new method -- the ``growing of turbulence''. 
We add to the equation a weak controlled pumping term and start adiabatic evolution of turbulence from statistically homogeneous Gaussian noise. 
After reaching a certain level of average intensity, we switch off the pumping and realize that the ``grown up'' turbulence is statistically stationary. 
We measure its Fourier spectrum, the probability density function (PDF) of intensity and the autocorrelation of intensity. 
Additionally, we show that, being adiabatic, our method produces stationary states of the integrable turbulence for the intermediate moments of pumping as well. 
Presently, we consider only the turbulence of relatively small level of nonlinearity; however, even this ``moderate'' turbulence is characterized by enhanced generation of rogue waves.
\end{abstract}

\maketitle

%-------------------------------------------------------------------------------------------------------------
%-------------------------------------------------------------------------------------------------------------

\section{Introduction}
\label{Sec:Intro}

The theory of integrable turbulence is one of the hottest topics in modern physics of nonlinear phenomena. 
The integrable turbulence is a state of an integrable system with infinite number degrees of freedom, such that many of them are exited in a random way. 
This state should be described statistically, and this is the subject of the integrable turbulence theory, the concept of which was introduced in 2009~\cite{zakharov2009turbulence} by one of the authors of the present paper.

The analytical approach to the theory of integrable turbulence is possible in two opposite situations:
\begin{enumerate}
 \item When the nonlinearity is weak. In this case one can use the expansion in powers of nonlinearity. This way was outlined in~\cite{zakharov2009turbulence}.
 \item When the turbulence can be treated as an ensemble of solitons. The kinetic theory of rarefied solitonic gas was suggested in~\cite{zakharov1971kinetic} and essentially improved in~\cite{el2005kinetic}. Later, wavefield statistical characteristics of rarified solitonic gas were studied in~\cite{pelinovsky2013two,pelinovsky2017kdv}.
\end{enumerate}
Yet a lot of many interesting types of integrable turbulence (the turbulence with intermediate level of nonlinearity, the theory of dense solitonic gas) remain out of limits of analytical theory and can be studied by implementation of massive numerical experiments only~\cite{walczak2015optical,agafontsev2015integrable,akhmediev2016breather,suret2016single,randoux2016nonlinear,agafontsev2016integrable,gelash2018strongly,gelash2019bound,agafontsev2020integrable}.

In the present paper we study the most important and popular integrable system described by the focusing one-dimensional nonlinear Schr{\"o}dinger equation (1D-NLSE) and suggest a new approach -- the ``growing of integrable turbulence''. 
We add to the 1D-NLSE a small adiabatic pumping term, making the waves of small amplitude unstable, and observe development of this instability starting with statistically space-homogeneous Gaussian noise. 
When the average intensity reaches a certain controlled level, we switch off the pumping and leave the ``grown up'' state to develop according to the conservative 1D-NLSE. 
We realize that this ``grown'' turbulence isn't only statistically homogeneous, but is statistically stationary as well. 
We examine the Fourier spectrum of this system, the probability density function (PDF) of intensity and the autocorrelation of intensity. 
Additionally, we verify that, being adiabatic, our method produces stationary states of the integrable turbulence for the intermediate moments of pumping as well.

Presently, we limit ourselves with the ``grown up'' turbulence of relatively small level of nonlinearity, which is characterized by the ratio of the potential energy (related to nonlinearity) to the kinetic one (related to dispersion) of around $1/5$. 
However, even for this case, the PDF of intensity has ``fat tail'', indicating enhanced generation of rogue waves. 
We will continue our numerical experiments in the future. 

The paper is organized as follows.
In the next Section we describe our numerical methods.
In Section~\ref{Sec:Pumping} we discuss the two scenarios of the pumping -- the linear and the nonlinear ones.
In Section~\ref{Sec:Results} we demonstrate our results.
The final Section contains conclusions.

%-------------------------------------------------------------------------------------------------------------
%-------------------------------------------------------------------------------------------------------------

\section{Numerical methods}
\label{Sec:NumMethods}

Without loss of generality, we examine statistics of solutions for the following system of equations,
\begin{eqnarray}
\psi(t=0,x) = A_{0}f(x),\quad \overline{|f|^{2}}=1,&&\label{NLSE-1}\\
\left\{\begin{array}{rlllc}
i\psi_{t} + \psi_{xx} + |\psi|^{2}\psi = i\,\hat{p}\,\psi, & \mbox{while} & \overline{|\psi|^{2}}<1,\\
i\psi_{t} + \psi_{xx} + |\psi|^{2}\psi = 0, & \mbox{for} & \overline{|\psi|^{2}}=1, \end{array}\right.
\label{NLSE-2}
\end{eqnarray}
where $t$ is time, $x$ is spatial coordinate, $\psi$ is the wavefield, $f(x)$ is the function describing statistics of the initial noise, $A_{0}\ll 1$ is the noise amplitude and $\hat{p}$ is the pumping operator (linear or nonlinear).
For the numerical study, we consider the periodic problem $x\in[-L/2, L/2]$ with a very large period, $L=256\pi$; the overline denotes spatial averaging over this period,
$$
\overline{|f|^{2}} = \frac{1}{L}\int_{-L/2}^{L/2}|f|^{2}\,dx.
$$

In the absence of the pumping term $\hat{p}=0$, Eq.~(\ref{NLSE-2}) is the 1D-NLSE of the focusing type, which conserves an infinite series of invariants~\cite{zakharov1972exact,novikov1984theory}.
The first three of these invariants are wave action (in our notations equals to the average intensity),
\begin{equation}\label{wave-action}
N = \overline{|\psi|^{2}} = \frac{1}{L}\int_{-L/2}^{L/2}|\psi|^{2}\,dx = \sum_{k}|\psi_{k}|^{2},
\end{equation}
momentum
\begin{equation}\label{momentum}
P = \frac{i}{2L}\int_{-L/2}^{L/2}(\psi_{x}^{*}\psi-\psi_{x}\psi^{*})\,dx = \sum_{k}k|\psi_{k}|^{2},
\end{equation}
and total energy
\begin{eqnarray}
&& E = H_{l} + H_{nl}, \label{energy-1}\\
&& H_{l} = \overline{|\psi_{x}|^{2}} = \frac{1}{L}\int_{-L/2}^{L/2}|\psi_{x}|^{2}\,dx = \sum_{k}k^{2}|\psi_{k}|^{2}, \label{energy-2}\\
&& H_{nl} = -\frac{\overline{|\psi|^{4}}}{2} = -\frac{1}{2L}\int_{-L/2}^{L/2}|\psi|^{4}\,dx. \label{energy-3}
\end{eqnarray}
Here $H_{l}$ is the kinetic energy, $H_{nl}$ is the potential energy, $k=2\pi m/L$ is the wavenumber, $m\in\mathbb{Z}$ is integer and $\psi_{k}$ is the Fourier-transformed wavefield,
$$
\psi_{k}(t) = \frac{1}{L}\int_{-L/2}^{L/2}\psi(t,x)\,e^{-ikx}\,dx.
$$

In the case of system~(\ref{NLSE-1})-(\ref{NLSE-2}), the invariants of the 1D-NLSE change with time until the wave action $N=\overline{|\psi|^{2}}$ reaches unity, and then remain constant for all later times.
For adiabatic turbulence growth from one state close to the stationary state of the integrable turbulence to another, we take very small pumping, such that the motion is governed primarily by the terms of the 1D-NLSE, and also start simulations from small noise, $A_{0}\ll 1$, so that at the start of the growth stage the dynamics is almost linear (and, in the absence of the pumping, the linear turbulence would be stationary).

For numerical simulations, we use the pseudo-spectral Runge-Kutta fourth-order method in adaptive grid, with the grid size $\Delta x$ set from the analysis of the Fourier spectrum of the solution, see~\cite{agafontsev2015integrable} for detail.
The time step $\Delta t$ changes with $\Delta x$ as $\Delta t = h\,\Delta x^{2}$, $h\le 0.1$, in order to avoid numerical instabilities.
We have checked that, after turning off the pumping, the first ten integrals of motion of the 1D-NLSE are conserved by our numerical scheme up to the relative errors from $10^{-10}$ (the first three invariants) to $10^{-6}$ (the tenth invariant) orders.

The initial conditions are taken as white noise with wide super-Gaussian Fourier spectrum,
\begin{eqnarray}
f(x) = \sum_{k}\bigg(\frac{C_{n}}{\theta L}\bigg)^{1/2}\,e^{-|k|^{n}/\theta^{n}+ikx+i\phi_{k}}, \label{IC}
\end{eqnarray}
with $n=32$ and $\theta=4$.
Here $n$ is the exponent defining the shape of the Fourier spectrum, $\theta$ is characteristic width in the $k$-space, $\phi_{k}$ are random phases for each $k$ and each realization of the initial conditions, $C_{n}=\pi\, 2^{1/n}/\Gamma_{1+1/n}$ is the normalization constant such that $\overline{|f|^{2}}=1$ (see e.g. Eq.~(25) in~\cite{agafontsev2015integrable}) and $\Gamma$ is Gamma-function.
The noise spectrum is wide, as its characteristic width is much larger than unity, $\theta\gg 1$.
Also, for $A_{0}=1$, the noise would have ratio of the potential energy to the kinetic one equal to
\begin{eqnarray}
\alpha=\frac{|\langle H_{nl}\rangle|}{\langle H_{l}\rangle}\approx \frac{\Gamma_{1+1/n}}{\Gamma_{1+3/n}}\times\frac{3\cdot 2^{2/n}}{\theta^{2}}\approx 0.2, \label{alpha}
\end{eqnarray}
see~\cite{agafontsev2020integrable}, that corresponds to weakly nonlinear wavefield.
Below we will use the potential-to-kinetic energy ratio $\alpha$ to estimate the nonlinearity level of the wavefield.

After turning off the pumping, we start measurement of the statistical functions, averaging them over the ensemble of $200$ random realizations of initial conditions.
We have checked that larger ensemble size does not change the results.
We examine the ensemble-averaged kinetic $\langle H_{l}(t)\rangle$ and potential $\langle H_{nl}(t)\rangle$ energies, the fourth-order moment of amplitude $\kappa_{4}=\langle\overline{|\psi|^{4}}\rangle/\langle\overline{|\psi|^{2}}\rangle^{2}$, the PDF $\mathcal{P}(I,t)$ of relative wave intensity $I=|\psi|^{2}/\langle\overline{|\psi|^{2}}\rangle$, the wave-action spectrum,
\begin{equation}\label{wave-action-spectrum}
S_{k}(t) = \langle|\psi_{k}|^{2}\rangle/\Delta k,
\end{equation}
where $\Delta k = 2\pi/L$ is the distance between neighbor harmonics, and the autocorrelation of the intensity,
\begin{equation}\label{g2}
g_{2}(x,t) = \frac{\langle \overline{|\psi(y+x,t)|^{2}\cdot |\psi(y,t)|^{2}}\rangle}{\langle \overline{|\psi(y,t)|^{2}}\rangle^{2}}.
\end{equation}
Here $\langle ...\rangle$ means averaging over the ensemble of initial conditions and, in the latter relation, the overline denotes spatial averaging over the $y$ coordinate.
Note that, at $x=0$, the autocorrelation equals to the fourth-order moment, $g_{2}(0,t)=\kappa_{4}(t)$, and at $x\to\infty$ it must approach to unity, $g_{2}(x,t)\to 1$.
For the wave-action spectrum and the PDF, we use normalization conditions $\int S_{k}\,dk = N$ and $\int \mathcal{P}(I)\,dI = 1$, respectively.
Below we will also compare our numerical results for the PDF with the exponential function,
\begin{equation}\label{Rayleigh}
\mathcal{P}_{R}(I) = e^{-I},
\end{equation}
describing the distribution of intensity for a superposition of a multitude of uncorrelated linear waves with random Fourier phases, see e.g.~\cite{nazarenko2011wave}.

%-------------------------------------------------------------------------------------------------------------
%-------------------------------------------------------------------------------------------------------------

\section{The pumping term}
\label{Sec:Pumping}

First, let us consider scenario of the linear pumping $\hat{p}=b$.
In this case, the wave action evolves as
\begin{eqnarray}
\frac{dN}{dt} = 2 b N,&& \label{wave-action-L1}\\
N=\left\{\begin{array}{rlllc}
A_{0}^{2}\,e^{2 b t} & \mbox{for} & t<-\ln A_{0}/b,\\
1 & \mbox{for} & t\ge-\ln A_{0}/b. \end{array}\right. \label{wave-action-L2}
\end{eqnarray}
The characteristic time scale $t_{L}$ due to the effect of dispersion is connected with the characteristic length scale $\ell_{0}$ describing the function $\psi$ as $t_{L}=\ell_{0}^2$, see e.g.~\cite{agrawal2001nonlinear}.
At the initial time, the length scale is inverse-proportional to the noise spectral width, $\ell_{0}\sim 1/\theta$, see Eq.~(\ref{IC}).
Our numerical experiments indicate that, at the final time, the wave-action spectrum has the same characteristic width in the $k$-space as the initial noise, $\delta k\simeq\theta$, so that we may assume $\ell_{0}\sim 1/\theta$ and $t_{L}\sim 1/\theta^2$ for all times.
The nonlinear time describing the characteristic time scale due to nonlinearity is inverse-proportional to the wave action (average intensity), $t_{NL}=1/N$.
The latter changes from $A_{0}^{2}$ at the initial time to $1$ at the final time.
Finally, the characteristic time scale due to the pumping term equals to $t_{P}\simeq 1/2b$, see Eq.~(\ref{wave-action-L2}).
Thus, we can reach both (i) the adiabatic regime of the pumping and (ii) the close to linear evolution at the start of the growth stage only if
\begin{eqnarray}
\frac{1}{\theta^{2}}\ll \frac{1}{A_{0}^{2}}\ll \frac{1}{2b} \quad\leftrightarrow\quad 2b\ll A_{0}^{2}\ll \theta^{2}. \label{puming-adiabatic-L}
\end{eqnarray}
For instance, if we start from the initial noise amplitude $A_{0}=10^{-2}$ and use the pumping coefficient $b\lesssim 10^{-6}$, then the required evolution time before turning off the pumping is $t_{pf}=-\ln A_{0}/b\approx 4.6\times 10^{6}$.
For the statistical study involving ensembles with hundreds of realizations of initial conditions, such evolution times are difficult to reach with the currently available numerical resources.

We can also choose the nonlinear pumping term, for instance, proportional to the wave action, $\hat{p}=c N$.
In the sense of dependence on the wave action, such a pumping is similar to the saturating pumping modeling the ultra fast fiber lasers~\cite{bale2012modeling}.
Then, the wave action evolves as
\begin{eqnarray}
\frac{dN}{dt} = 2 c N^{2},&& \label{wave-action-NL1}\\
N=\left\{\begin{array}{rlllc}
\frac{A_{0}^{2}}{1-2 c A_{0}^{2}t} & \mbox{for} & t< \frac{1-A_{0}^{2}}{2 c A_{0}^{2}},\\
1 & \mbox{for} & t\ge \frac{1-A_{0}^{2}}{2 c A_{0}^{2}}, \end{array}\right. \label{wave-action-NL2}
\end{eqnarray}
and the characteristic time scale due to the pumping effect is inverse-proportional to the initial intensity, $t_{P}\simeq 1/2 c A_{0}^{2}$.
This leads to a different set of relations necessary for both (i) the adiabatic regime of the pumping and (ii) the close to linear evolution at the start of the growth stage,
\begin{eqnarray}
\frac{1}{\theta^{2}}\ll \frac{1}{A_{0}^{2}}\ll \frac{1}{2 c A_{0}^{2}} \quad\leftrightarrow\quad 2c\ll 1 \,\,\mbox{and}\,\, A_{0}^{2}\ll \theta^{2}. \label{puming-adiabatic-NL}
\end{eqnarray}
Thus, if we use parameters $c=A_{0}=10^{-2}$ similar to those for the described above linear pumping case, the required evolution time before turning off the pumping $t_{pf}=(1-A_{0}^{2})/2 c A_{0}^{2}\approx 5\times 10^{5}$ turns out to be one order of magnitude smaller.

There is also another advantage of the nonlinear pumping, that is especially valuable in combination with numerical schemes utilizing adaptive grids.
Specifically, the appearance of large gradients that require usage of fine discretization is expected mostly when the wave action (average intensity) reaches unity order, $N\sim 1$; such a behavior is confirmed experimentally by the performance of our numerical scheme.
For the linear pumping scenario, the system spends in evolution from $N=1/2$ to $N=1$ the time $\delta t_{1/2}=\ln 2/2b$, as can be easily calculated from Eq.~(\ref{wave-action-L2}).
For the considered above parameters $A_{0}=10^{-2}$ and $b=10^{-6}$, it equals to $\delta t_{1/2}\approx 3.5\times 10^{5}$.
For the nonlinear pumping with $c=A_{0}=10^{-2}$, the corresponding time $\delta t_{1/2}=1/2c = 50$, i.e., four orders of magnitude smaller.

Thus, for the nonlinear pumping, the system spends most of its evolution having very small wave action, when the adaptive numerical scheme resolves the wavefield accurately using comparatively small number of points $\mathcal{M}$.
Simulation of unit evolution time $\delta t=1$ with our method requires $\mathcal{O}(\mathcal{M}^{3}\log\mathcal{M})$ operations -- $\mathcal{O}(\mathcal{M}\log\mathcal{M})$ for the FFT multiplied by $\mathcal{O}(\mathcal{M}^{2})$ time steps -- that results in huge advantage in the overall simulation time compared to the linear pumping scenario.

For this reason, in the present paper we use only the nonlinear pumping term proportional to the wave action, $\hat{p}=c N$, and determine the evolution time for the growth stage via relation~(\ref{wave-action-NL2}).
However, we have checked that usage of the linear pumping term leads to qualitatively the same results for the statistical functions describing the integrable turbulence after turning off the pumping.

\begin{figure*}[t]\centering
\includegraphics[width=8.5cm]{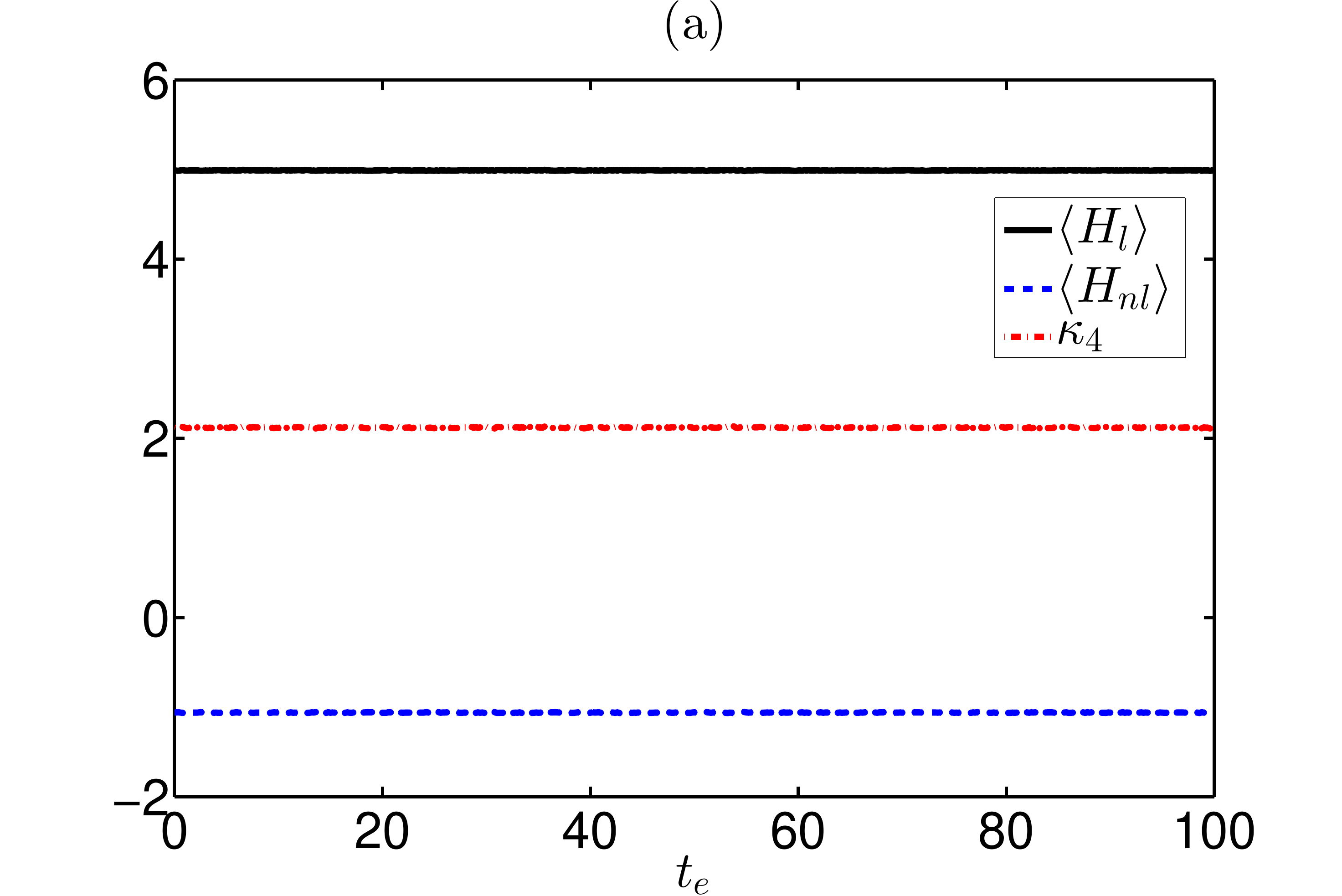}
\includegraphics[width=8.5cm]{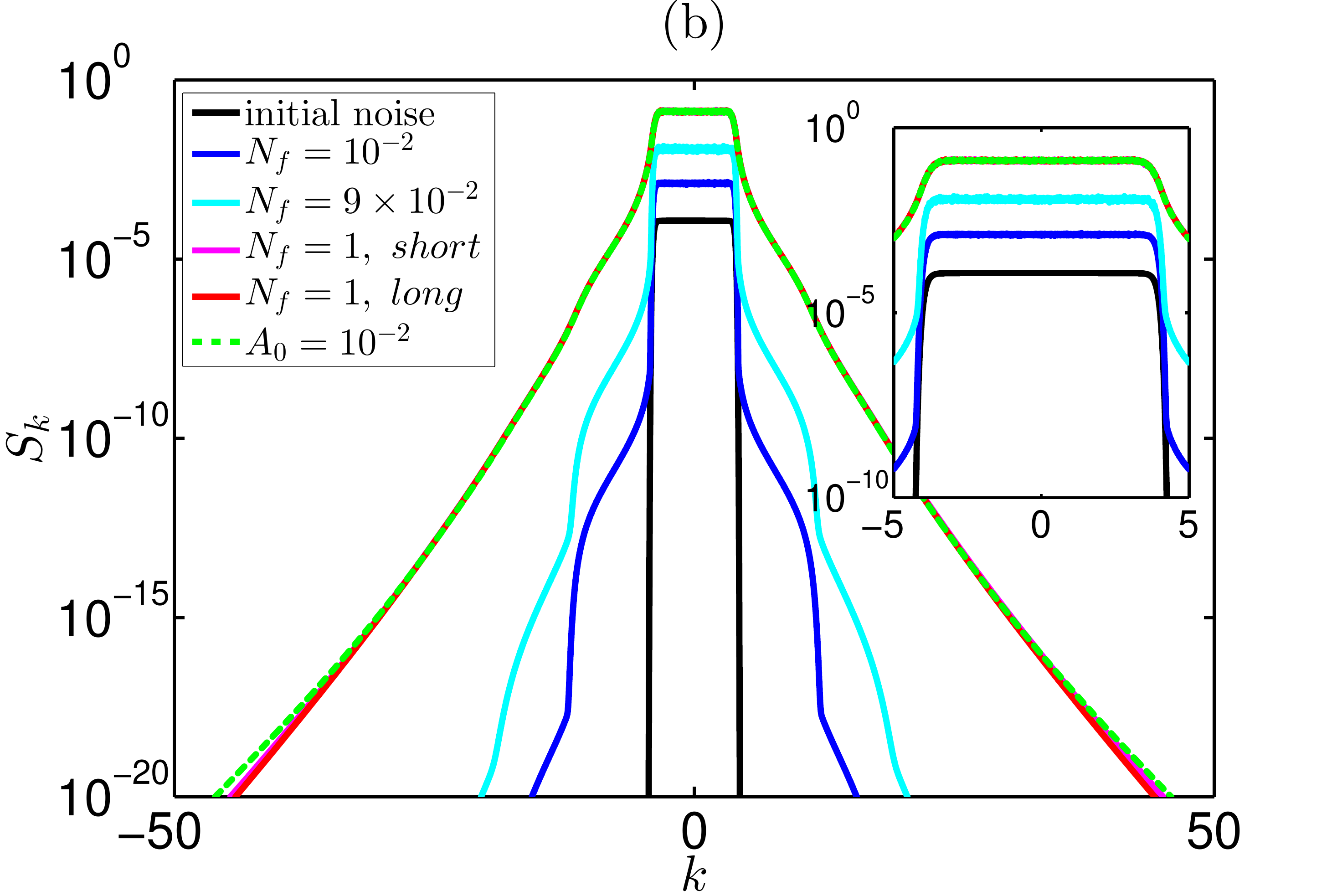}\\
\includegraphics[width=8.5cm]{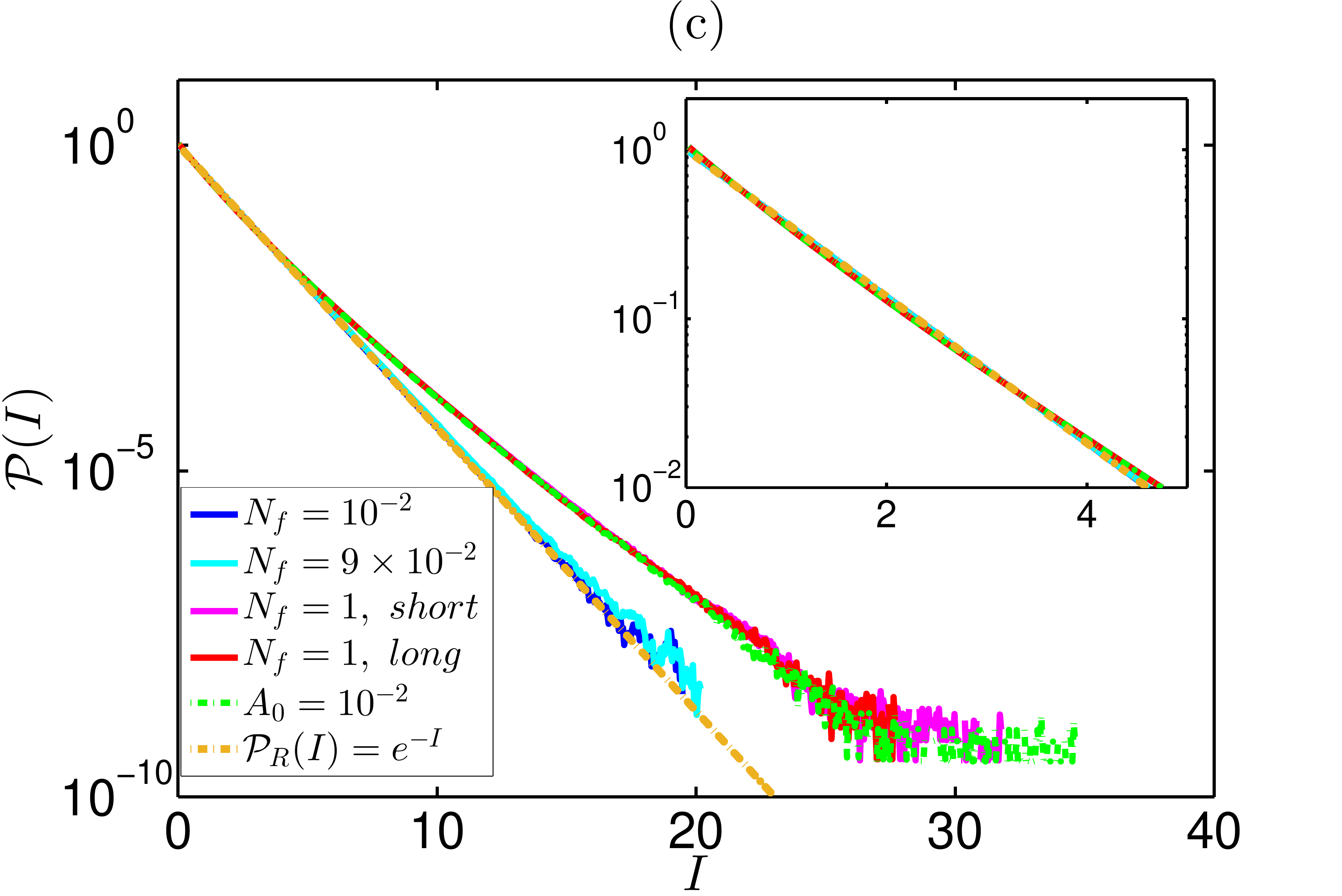}
\includegraphics[width=8.5cm]{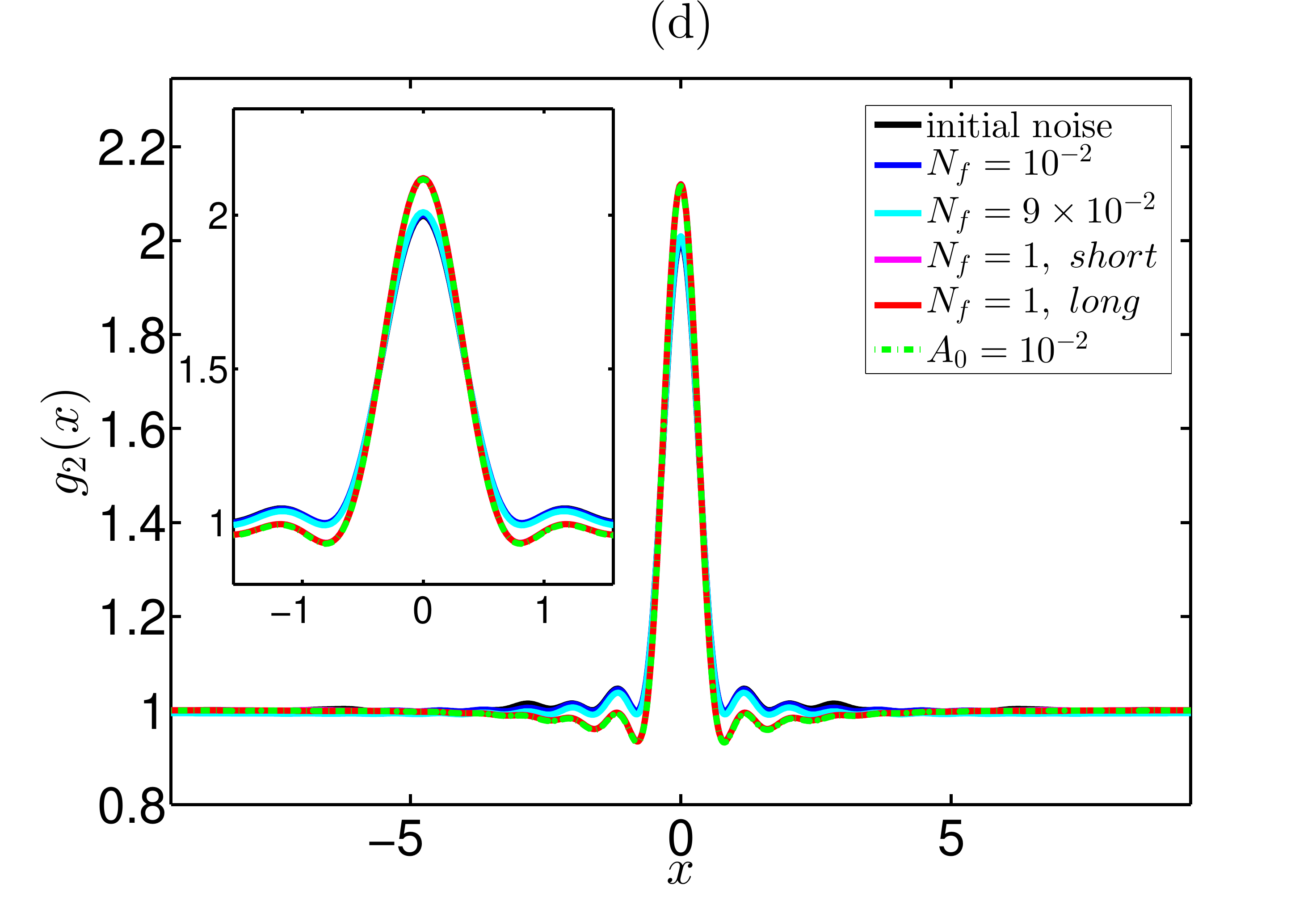}

\caption{\small {\it (Color on-line)}
(a) Ensemble-averaged kinetic energy $\langle H_{l}\rangle$, potential energy $\langle H_{nl}\rangle$ and the fourth-order moment $\kappa_{4}$ versus time $t_{e}=t-t_{pf}$, where $t_{pf}$ is the moment when the pumping is turned off; the initial noise amplitude is $A_{0}=3\times 10^{-2}$, the final average intensity is unity, $N_{f}=1$.
(b-d) Averaged over ensemble and time statistical functions of the integrable turbulence after the pumping is turned-off: (b) the wave-action spectrum $S_{k}$, (c) the PDF $\mathcal{P}(I)$ of relative wave intensity $I=|\psi|^{2}/\langle\overline{|\psi|^{2}}\rangle$ and (d) the autocorrelation of intensity $g_{2}(x)$.
In figures (b-d), all lines except for the green dashed lines correspond to the experiment with the initial noise amplitude $A_{0}=3\times 10^{-2}$.
In particular, the black lines mark the statistical functions for the initial noise, the blue lines indicate the interrupted growth stage with the final intensity $N_{f}=10^{-2}$ and time averaging in $t_{e}\in [0, 20]$, the cyan -- $N_{f}=9\times 10^{-2}$ with $t_{e}\in [0, 20]$, the pink -- $N_{f}=1$ with $t_{e}\in [0, 20]$ and the red -- $N_{f}=1$ with $t_{e}\in [80, 100]$.
The green dashed lines show the experiment with the initial noise amplitude $A_{0}=10^{-2}$ and the final intensity $N_{f}=1$, with the results averaged over time $t_{e}\in [0, 20]$.
The insets in panels (b-d) show the same functions as in the main figures with smaller scales, and the brown dash-dot line in panel (c) indicates the exponential PDF~(\ref{Rayleigh}).
The PDF for the initial noise is not shown, as it coincides (by construction) with the exponential PDF.
}
\label{fig:fig1}
\end{figure*}

%-------------------------------------------------------------------------------------------------------------
%-------------------------------------------------------------------------------------------------------------

\section{Results}
\label{Sec:Results}

In order to confirm the adiabatic process of turbulence growth from one state close to the stationary state of the integrable turbulence to another, we perform several numerical experiments.

In the first experiment, we start from the initial noise with amplitude $A_{0}=3\times 10^{-2}$, use nonlinear pumping term $\hat{p}=c N$ with $c=10^{-2}$, wait until the wave action reaches unity, turn off the pumping and measure the statistical functions.
Figure~\ref{fig:fig1}(a) demonstrates the ensemble-averaged kinetic $\langle H_{l}\rangle$ and potential $\langle H_{nl}\rangle$ energies and the fourth-order moment of amplitude $\kappa_{4}$ versus time $t_{e}=t-t_{pf}$ after the moment $t_{pf}$ when the pumping was turned off.
As shown in the figure, the three functions do not change with time for sufficiently long evolution, so that at $t_{e}=0$ the turbulence can already be considered as very close to stationary.
For $t_{e}>0$, the potential-to-kinetic energy ratio turns out to be $\alpha=|\langle H_{nl}\rangle|/\langle H_{l}\rangle\approx 0.212$, i.e. only about $6$\% larger than for the initial white noise scaled to unit average intensity, see Eq.~(\ref{alpha}); the small value of $\alpha$ indicates that the turbulence is weakly nonlinear.
For $t_{e}>0$, the fourth-order moment equals to $\kappa_{4}\approx 2.12$, that is slightly larger than the value of $2$ characterizing a superposition of a multitude of uncorrelated linear waves with random phases.
The latter hints that the rogue waves are generated slightly more frequently than for a purely linear system.

To ensure that, after turning off the pumping, the integrable turbulence is very close to stationary, we follow~\cite{agafontsev2020integrable} and compare the statistical functions averaged over ensemble of initial conditions and two different time intervals $t\in[0,20]$ (indicated with magenta lines in Fig.~\ref{fig:fig1}(b-d)) and $t\in[80,100]$ (red lines) -- for the wave-action spectrum $S_{k}$, the PDF $\mathcal{P}(I)$ of relative wave intensity $I=|\psi|^{2}/\langle\overline{|\psi|^{2}}\rangle$ and the autocorrelation of intensity $g_{2}(x)$.
As shown in the figures, the results are identical, so that after turning off the pumping the integrable turbulence can be considered stationary. 

In Fig.~\ref{fig:fig1}(b-d), the magenta and the red lines also coincide with the dashed green line, which indicates results for a different numerical experiment with the same parameters as for the first experiment, except for the smaller initial noise amplitude $A_{0}=10^{-2}$.
The match of the results confirms that the initial noise for the first experiment $A_{0}=3\times 10^{-2}$ is small enough to seed the adiabatically growing integrable turbulence.

We now describe the basic features of statistical functions shown in Fig.~\ref{fig:fig1}(b-d) for the first experiment (magenta and red lines).
In particular, the wave-action spectrum inherits the flat profile of the noise spectrum at small and moderate wavenumbers $|k|\le 4$, and decays slightly slower than exponential at large wavenumbers $|k|\gtrsim 10$, Fig.~\ref{fig:fig1}(b).
The PDF deviates from the exponential function~(\ref{Rayleigh}), exceeding it significantly at large intensities, Fig.~\ref{fig:fig1}(c); for $I=20$, the excess reaches about $2$ orders of magnitude.
Together with the slightly elevated fourth-order moment $\kappa_{4}\approx 2.12$ compared with the value of $2$ characterizing a superposition of a multitude of uncorrelated linear waves, this is a sign of enhanced appearance of rogue waves.
The autocorrelation of intensity is a bell-shaped function at small distances $|x|\lesssim 1$ with the maximum slightly larger than $2$, $\max g_{2}(x)=g_{2}(0)=\kappa_{4}\approx 2.12$, and is nearly indistinguishable from unity at larger distances $|x|\gtrsim 4$.

As we have shown, when the wave action reaches unity and we turn off the pumping, the resulting integrable turbulence is practically stationary.
To confirm that, during the growth stage, the turbulence goes through the similar almost-stationary states defined by the current set of the (very slowly changing) integrals of motion, we perform two more experiments, in which we turn off the pumping earlier.
Namely, in the first of these experiments we turn off the pumping when the wave action reaches $N_{f}=10^{-2}$ (blue lines in Fig.~\ref{fig:fig1}(b-d), time averaging over $t_{e}\in[0,20]$), and in the second -- at $N_{f}=9\times 10^{-2}$ (cyan lines, $t_{e}\in[0,20]$).
Repeating the procedure described above, for each of these two experiments we have compared the statistical functions averaged over the ensemble of initial conditions and two different time intervals $t_{e}\in[0,20]$ and $t_{e}\in[80,100]$, and found no difference (the curves corresponding to $t_{e}\in[80,100]$ are not shown in the figure for better visibility).
Hence, we can conclude that, during the growth stage, the intermediate states are also very close to the stationary states of the integrable turbulence.

Note that for the additional experiment with $N_{f}=10^{-2}$, after turning off the pumping, the potential-to-kinetic energy ratio and the fourth-order moment equal to $\alpha\approx 2.02\times 10^{-3}$ and $\kappa_{4}\approx 2.001$, while for the experiment with $N_{f}=9\times 10^{-2}$ -- to $\alpha\approx 1.83\times 10^{-2}$ and $\kappa_{4}\approx 2.008$.
Hence, even though at these intermediate states the turbulence is almost linear, the fourth-order moment indicates increasing deviation from Gaussian statistics.

For the two intermediate states with $N_{f}=10^{-2}$ and $N_{f}=9\times 10^{-2}$, the PDF and the autocorrelation of intensity shown in Fig.~\ref{fig:fig1}(c,d) almost coincide with the exponential PDF~(\ref{Rayleigh}) and the autocorrelation of intensity for the initial noise, respectively.
The most significant change with the final intensity $N_{f}$ is observed for the wave-action spectrum -- compare the black (initial noise), blue ($N_{f}=10^{-2}$), cyan ($N_{f}=9\times 10^{-2}$) and pink or red ($N_{f}=1$) lines in Fig.~\ref{fig:fig1}(b).
While the flat profile at small and moderate wavenumbers $|k|\le 4$ goes up with increasing final intensity (we remind that the wave-action spectrum is normalized to intensity), at large wavenumbers the spectrum acquires tails that decay nontrivially with the wavenumber and widen with increasing $N_{f}$.
At $N_{f}=1$, the turbulence is weakly nonlinear, $\alpha\approx 0.212$, and the tails decay slightly slower than exponentially.

From the results presented above, we can conclude that the integrable turbulence can be grown adiabatically from a small noise by a temporary addition of a small pumping term.
During this process, the turbulence goes consequentially through states, that a very close to the stationary states of the integrable turbulence defined by the current set of the (slowly changing) integrals of motion.

%-------------------------------------------------------------------------------------------------------------
%-------------------------------------------------------------------------------------------------------------

\section{Conclusions}
\label{Sec:Conclusions}

In the present paper we have suggested a new approach to the studies of integrable turbulence, that consists in adiabatic growing of turbulence from small noise by a temporary addition of a small pumping term to the core integrable equation.
The small level of the initial noise ensures the close to linear evolution of the system at the start of the growth stage, thus making the turbulence at this time almost stationary.
The usage of a small pumping term, such that its influence is much smaller than that of all other terms, allows us to grow the turbulence adiabatically, i.e., when the dynamics is defined mostly by the core integrable equation and the main role of the pumping is reduced to the slow change of the integrals of motion.
In combination, this design allows the adiabatic process of turbulence growth from one state very close to the stationary state of the integrable turbulence to another, with the intermediate states defined by the current set of the integrals of motion.

We have performed a numerical experiment designed according to these principles and confirmed the described above behavior.
As a seed for the turbulence growth, we have used a wide-spectrum noise, that led us to weakly nonlinear turbulence after the finish of the growth stage.
Our motivation was two-fold: first, noise in nature has typically wide spectrum and, second, our preliminary simulations have shown that narrower initial noise requires more simulation time.
The same preliminary experiments demonstrate the dependency: the narrower the noise spectrum, the larger the nonlinearity of the resulting integrable turbulence.
We will continue this line of study in the future publications.

Nevertheless, even the resulting weakly nonlinear turbulence is characterized by the heavy-tailed PDFs of relative wave intensity and elevated value of the fourth-order moment $\kappa_{4}>2$, that indicate enhanced generation of rogue waves.
The wave-action spectrum inherits the profile of the noise spectrum at small and moderate wavenumbers, and decays slightly slower than exponential at large wavenumbers.
The autocorrelation of intensity turns out to be a bell-shaped function at small distances, and quickly converges to unity at larger distances.

Until now, the studies of integrable turbulence were focused on examination of specific initial conditions, such as the condensate~\cite{agafontsev2015integrable,kraych2019statistical}, the cnoidal wave~\cite{agafontsev2016integrable}, the partially coherent wave~\cite{walczak2015optical,suret2016single,agafontsev2020integrable} and its superposition with the condensate~\cite{akhmediev2016breather}, and also the soliton gas~\cite{costa2014soliton,gelash2018strongly,gelash2019bound,redor2019experimental}.
In all such studies, it was implicitly assumed that the initial conditions were somehow prepared by an external actor, that resembles a setting of a laboratory experiment.
We believe that our approach of adiabatically growing integrable turbulence is very promising, as it accounts explicitly for generation of the initial conditions and may model processes in nature more accurately.
We will continue our studies with this approach in the near future.

%-------------------------------------------------------------------------------------------------------------
%-------------------------------------------------------------------------------------------------------------

\begin{center}
\textbf{Acknowledgements}
\end{center}

The authors thank A.\,A.~Gelash for fruitful discussions.
Simulations were performed at the Novosibirsk Supercomputer Center (NSU).
The work of both authors was supported by the Russian Science Foundation Grant No. 19-72-30028.

%-------------------------------------------------------------------------------------------------------------
%-------------------------------------------------------------------------------------------------------------

\bibliographystyle{apsrev4-1}
\bibliography{refs}

\end{document}